# THE CHEMISTRY OF IRREGULAR GALAXIES[2]

Don Garnett[1]

Astronomy Department, University of Minnesota

116 Church St., S. E., Minneapolis, MN 55455 USA

## RESUMEN

I review the current data on abundances in irregular galaxies, with special emphasis on recent results on C/O and Si/O from observations with the Hubble Space Telescope. The abundance data for He, C, N, and O in irregular galaxies is discussed in the context of models for the evolution of dwarf galaxies.

## ABSTRACT

I review the current data on abundances in irregular galaxies, with special emphasis on recent results on C/O and Si/O from observations with the Hubble Space Telescope. The abundance data for He, C, N, and O in irregular galaxies is discussed in the context of models for the evolution of dwarf galaxies.

## 1. INTRODUCTION

Irregular galaxies continue to be fascinating objects for the study of the evolution of galaxies, and for comparison with spiral galaxies. Irregulars are typically low-mass, gas rich, and blue. The star formation activity in irregulars is not organized by spiral density waves compressing the gas, and yet many have prolific or "violent" star formation activity! The rotation of irregular galaxies is characterized by solid body rotation as opposed to the differential rotation observed in spirals. This suggests that differential shear does not operate in irregulars to redistribute gas and heavy elements within them, and leads one to hope that irregulars may be simpler systems to study than spirals. Irregulars also have low heavy element abundances, with O/H typically < 25% of solar, suggesting that irregulars may be "less evolved" than spirals. Irregulars thus provide opportunities to study star formation and evolution processes in environments very different from the solar neighborhood and spirals, and to compare abundances in metal-poor galaxies with the halo of the Milky Way, to see if chemical evolution has proceed in similar ways. Abundance measurements in very low metallicity irregulars provide a benchmark to compare with measurements in high-redshift galaxies and QSO absorption-line systems; and measurements of helium abundances in irregulars have given the most precise determinations of the primordial He fraction, helping constrain theoretical models for primordial big bang nucleosynthesis and cosmological parameters.

The last review of abundances in irregular galaxies for Tex-Mex was done at the original Tex-Mex meeting by Dufour (1986). Since then, a great deal of observational material has been obtained, including the first results from the Hubble Space Telescope, so it seemed a propitious time to have another review. I will first review what we know about abundances in irregulars from ground-based studies, and then discuss recent results on carbon and silicon in irregulars from UV spectroscopy with HST. Finally, I will discuss briefly the abundance results in the light of models for the evolution of irregular galaxies.

---

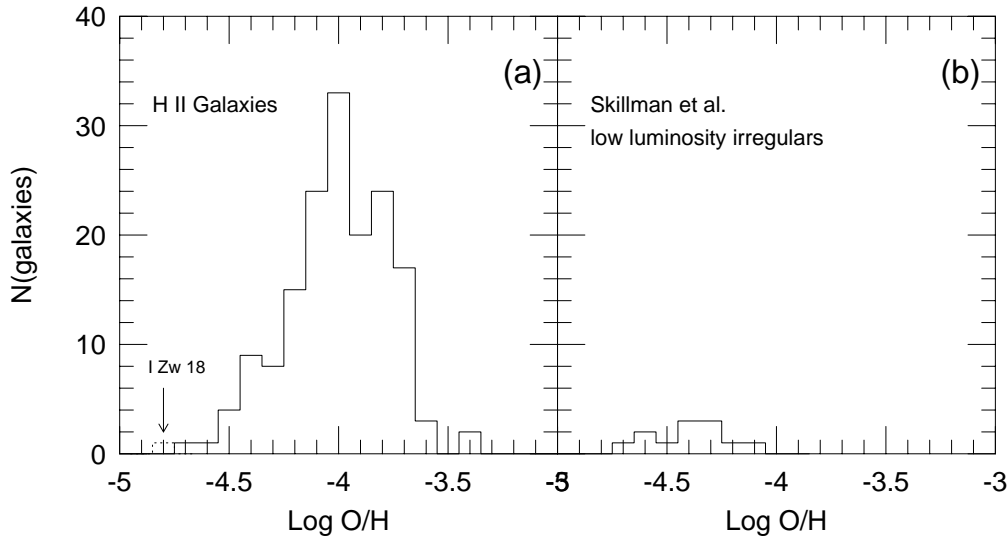

Fig. 1.— Histogram of distribution of O/H in irregular galaxies. Left: Data for H II galaxies from Peña et al. (1991) and Masegosa et al. (1994). Right: data for low luminosity galaxies from Skillman et al.

## 2. GROUND-BASED ABUNDANCE MEASUREMENTS IN IRREGULAR GALAXIES

Element abundances in irregular galaxies are most often determined from measurements of collisionally-excited and recombination emission lines emitted by H II regions in the galaxies. The earliest such studies of abundances in the Magellanic Clouds and irregular galaxies (Peimbert & Spinrad 1970, Searle & Sargent 1972, Peimbert & Torres-Peimbert 1974, 1976, Dufour 1975) demonstrated that these galaxies have sub-solar abundances of O and N, but that He was essentially normal. This provided striking confirmation that He is mainly product of Big Bang nucleosynthesis. At the same time, Searle & Sargent noted that the low metallicities and very blue colors of dwarf irregular galaxies indicated that either (1) dwarf galaxies are very young, or (2) they experience episodic star formation ("starbursts") separated by long quiescent periods. Subsequently, the first spectroscopic studies of large samples of dwarf irregulars in the late 1970s and early 1980s (Lequeux et al. 1979, French 1980, Talent 1980, Kinman & Davidson 1981, Kunth & Sargent 1983, Hunter, Gallagher, & Rautenkrantz 1982) confirmed that irregulars typically have low metallicities, O/H from 2% to 25% of solar, and that $<N/O> \approx 0.03$, only about 1/3 the solar value. They also provided the first reliable extrapolations of Y to zero O/H, showing that the primordial He fraction $Y_p$ is close to standard Big Bang nucleosynthesis predictions.

Since Dufour's review there have been a number of studies devoted to obtaining high quality spectroscopy of H II regions in irregulars, particularly with regard to improving the precision in $Y_p$ (e.g., Pagel et al 1992, Terlevich et al. 1991 Spectrophotometric Catalogue of H II Galaxies; Skillman et al. 1989a et seq.). Figure 1a shows the abundance distribution of H II galaxies from the combined samples of Peña et al. (1991) and Masegosa et al. (1994). The histogram shows that the H II galaxies discovered from objective-prism surveys are predominately metal-poor, with a distribution strongly peaked around log O/H $\approx -4$, similar to the distribution shown in Dufour's (1986) Figure 1. Note that the larger samples have filled in the tail of the distribution between I Zw 18 and the bulk of the sample at log O/H $> -4.5$, but still no galaxy has been found with O/H smaller than in I Zw 18. This abundance distribution raises some still-unanswered questions: is the lack of very metal-poor galaxies a selection effect due to the difficulty of measuring the increasingly weak forbidden lines in these H II regions? Or are very metal-poor galaxies simply very rare - because the luminosity function for dwarf galaxies is falling off rapidly at low luminosities? Or are very metal-poor H II galaxies rare because they are starbursting with a much longer duty cycle? Skillman et al. (1989a,b, 1994) confirmed that irregular galaxies exhibit a correlation between metallicity and galaxy luminosity, and showed that a directed study of very low-luminosity systems could efficiently discover very metal-poor galaxies (Fig. 1b). The Sloan Digital Sky Survey, looking at both magnitudes and redshifts for a very large galaxy sample, could provide sufficient data on low-luminosity H II galaxies to answer some of these questions. By contrast, the sharp cutoff for log O/H $> -3.7$ is probably a selection effect due to the difficulty of measuring electron temperatures at higher abundances.

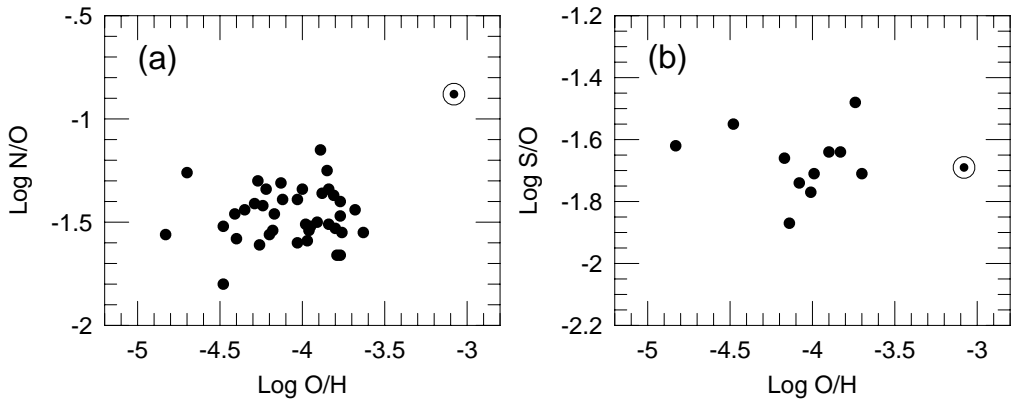

Fig. 2. (a) N/O vs. O/H for irregular galaxies, adapted from Garnett (1990). (b) S/O vs. O/H for irregular galaxies; after Garnett (1989).

At the same time, these studies have improved our understanding of the behavior of other elements in irregular galaxies. As noted already, N/O was shown to be constant with O/H in irregulars, and low with respect to the Galaxy (Figure 2a). This has reinforced the idea that nitrogen has a primary nucleosynthesis component that dominates in low-metallicity systems (Matteucci & Tosi 1985). Garnett (1989) showed that S/O is constant at the solar ratio in irregulars (Figure 2b), while Peña et al. (1991) and Masegosa et al. (1994) determined that Ne/O is essentially constant at the solar ratio. These results suggest that the products of massive star nucleosynthesis vary in lockstep, and thus neither the yields nor the massive star IMF are varying significantly with metallicity.

## 3. UV SPECTROSCOPY OF H II REGIONS WITH HST

Now I would like to present some recent results from UV spectroscopy of H II regions with the *Hubble Space Telescope*. The UV region from 1600-2000 Å contains a very nice series of collisionally-excited intercombination transitions from O III] (1661, 1666 Å), N III] (1748-1754 Å), Ne III] (1814 Å), Si III] (1883, 1892 Å), and C III] (1906, 1909 Å). These species have similar ionization potentials and the transitions similar excitation energies, so one can use these lines to determine abundance ratios for C, N, O, Si, and Ne with greatly diminished uncertainties due to errors in ionization and $T_e$; uncertainties from reddening are also reduced compared to abundances from UV/optical line ratios.

C and Si are important for a variety of problems in astronomy. The formation and evolution of CO in the ISM must be regulated by relative abundances of C and O at some level, and so one would like to know how both C and O evolve with time in galaxies. C and Si are both constituents of interstellar grains (Mathis 1990); the abundances of C and Si in the ISM will therefore reflect on the dust-to-gas ratio in the ISM. Moreover, study of the abundances of refractory elements in H II regions provides information on the ability of grains to survive in the ionized gas near hot stars. Since O is synthesized almost entirely in massive stars, while C is produced in both massive and intermediate mass stars, some C is ejected with a time delay with respect to O so a comparison of the evolution of C, N, and O offers potential insight into the star formation history of galaxies. Finally, Steigman et al. (1989) suggested that C may be a better measure of stellar He production than O, and so the C/H vs. Y relation may offer a more reliable determination of $Y_p$.

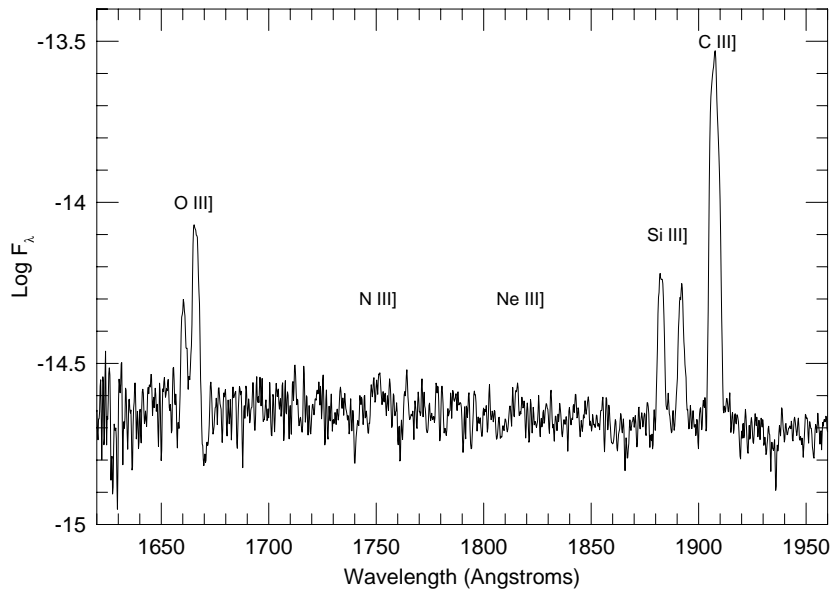

Fig. 3.— FOS spectrum of the H II region N88A in the SMC. The positions of several important intercombination transitions from C III, N III, O III, Ne III, and Si III are marked.

Little information on C and Si abundances has been available until now; both elements have no easily measured transitions in the optical spectrum, and require space-based observing. *IUE* provided some information on C III], but the carbon abundances derived from those data had quite large uncertainties (Dufour et al. 1984). Therefore, we set out to determine improved abundances for these elements using *HST*. We obtained UV spectra of seven H II regions in irregular galaxies during Cycles 2 and 3, using the FOS with G190H and a 1″ aperture to cover the 1600-2300 Å region; we set out specifically to measure the C III], Si III], *and* O III] emission lines *simultaneously*, to reduce the uncertainties in our derived C/O and Si/O values. Figure 3 shows a sample spectrum of the 1600 - 2000 Å spectral region.

Our results are shown in Figures 4 and 5 (Garnett et al. 1994, 1995). We find that the C/O ratio increases monotonically as a function of O/H, from about 1/4 solar at the lowest metallicities up to solar values. At the same time, C/N also appears to increase with O/H in the irregular galaxies, but the trend appears to be broken between the irregulars and the solar neighborhood. This seems to indicate that most of the nitrogen production in our Galaxy is decoupled from that of carbon and oxygen.

One possible interpretation of the trend in C/O is that the most metal-poor galaxies are young and dominated by the products of early enrichment by massive stars, while more metal-rich galaxies show increasing, delayed contributions of carbon from intermediate-mass stars. However, if C and N both come primarily from intermediate-mass stars, it is difficult to reconcile the steadily increasing C/O with the break in C/N between the LMC and the Galaxy. This leads one to suspect that carbon in the irregular galaxies has been produced mainly in massive stars. The increasing C/O with metallicity could then be the result of an increasing yield of carbon from massive stars, relative to oxygen, due to metallicity-dependent stellar mass loss. Maeder (1992) has shown that mass loss in stellar winds, on the main sequence and during the LBV and Wolf-Rayet phases, can have a very significant effect on the evolution and nucleosynthetic yields from massive stars, particularly for He, C, and O. If stellar winds are driven by radiation pressure, the mass loss rates should depend on atmospheric opacity and thus on metallicity; measurements of WR/O and WN/WC star ratios in regions of different metallicity indicate that this is probably true (Maeder 1991). We have compared our results for C/O with chemical evolution calculations by Carigi (1994) and Prantzos et al. (1994), which use Maeder's yields; these models reproduce the observed trend in C/O better than the model of Matteucci & François (1989) based on the massive star yields of Woosley & Weaver (1986) (without stellar mass loss); in fact, our C/O data suggest that mass loss rates might be even higher than those used by Maeder. However, more data on C/O in galaxies are needed to delineate the trend in C/O more clearly before definite conclusions can be made.

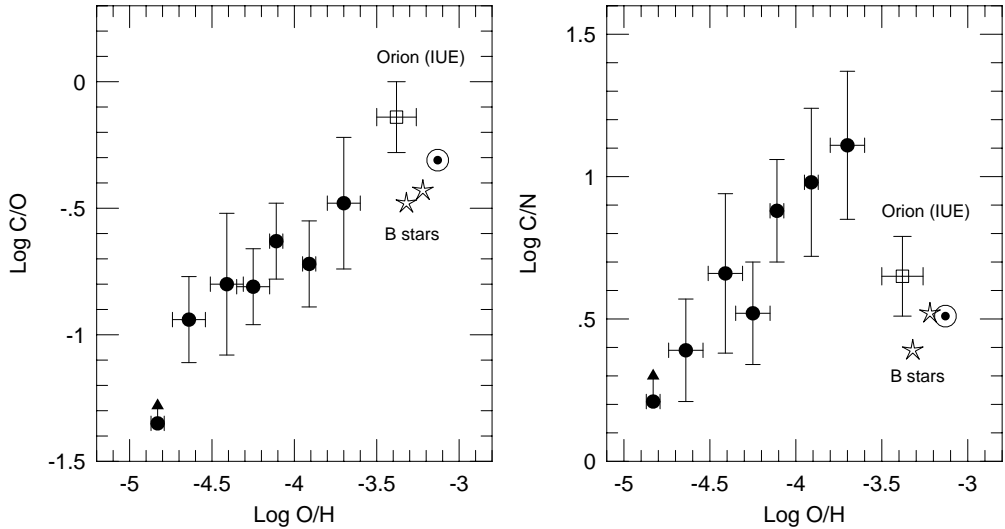

Fig. 4. Left: C/O vs. O/H for metal-poor irregular galaxies from Garnett et al. (1994) (filled circles). Open square is the Orion Nebula from Walter, Dufour & Hester (1992); stars are solar neighborhood B stars from Gies & Lambert (1992) and Cunha & Lambert (1994); solar value is from Grevesse & Noels (1993). Right: C/N vs. O/H; symbols are the same as for the left panel.

Figure 5 shows that Si/O appears to be remarkably constant, and not very different from values observed in the solar neighborhood; the mean log Si/O = $-1.59\pm0.07$, about 60% of the solar ratio. This result suggests that the depletion of Si onto dust grains is small in giant H II regions compared to dense neutral interstellar gas (Sofia, Cardelli, & Savage 1994), and that grain modification may be occurring in the H II region environment. However, measurements of Fe abundances in Galactic H II regions (e.g., Peimbert et al. 1993) indicate that 90% of the Fe in H II regions is still in grains, so grain destruction is quite incomplete. Sofia et al. (1994) have shown that the Fe and Mg in grain cores is not mostly in silicates, as once supposed, and that Si is predominantly in grain mantles. Therefore some process seems to erode Si from grain mantles but leave the Fe-Mg cores intact inside the ionized gas. Some possibilities include shocks driven by winds from O and B stars, or the shock which precedes the D-type ionization front. Future measurements of refractory element abundances in H II regions should provide exciting new information on the physical properties of grains in the vicinity of hot stars.

### 4. CHEMISTRY AND EVOLUTION OF IRREGULAR GALAXIES

The surprising blue colors and low abundances of I Zw 18 and II Zw 40 led Searle & Sargent (1972) to propose that galaxies like these were either extremely young or undergoing intermittent starbursts. Since 1986, there have been a number of new developments with regard to both models for dwarf galaxies, which I discuss below.

*Primordial Galaxies?*

The utter lack of known galaxies with abundances smaller than in I Zw 18, despite concerted observational effort, has been a puzzle. This led Kunth & Sargent (1986) to suggest that I Zw 18 was indeed a primordial object, in which the H II regions we observe have been self-enriched by supernovae from the current starburst; they predicted that neutral gas associated with I Zw 18 should be essentially pristine. Kunth et al. (1994) obtained *HST* measurements of interstellar O I $\lambda1302$, Ly-$\alpha$ absorption in I Zw 18 against the OB star continuum of the ionizing cluster, used the H I 21cm emission line data of Viallefond et al. (1987) to estimate the velocity dispersion in the neutral gas, and determined that O/H ≈ 1/1000 solar in the neutral gas, whereas O/H ≈ 1/50 solar in H II region. However, Pettini & Lipman (1995) have criticized that analysis: they argue that the O I line is highly saturated, and that the b-value derived from H I 21cm *emission* is not likely to be appropriate for the *absorption* line of sight. They modeled the Kunth et al. O I line for various values of b and found that for reasonable values of the velocity dispersion they could derive O/H anywhere between solar and 0.001 solar. Pending observations of unsaturated interstellar lines in I Zw 18, the question of abundances in the neutral material remains open.

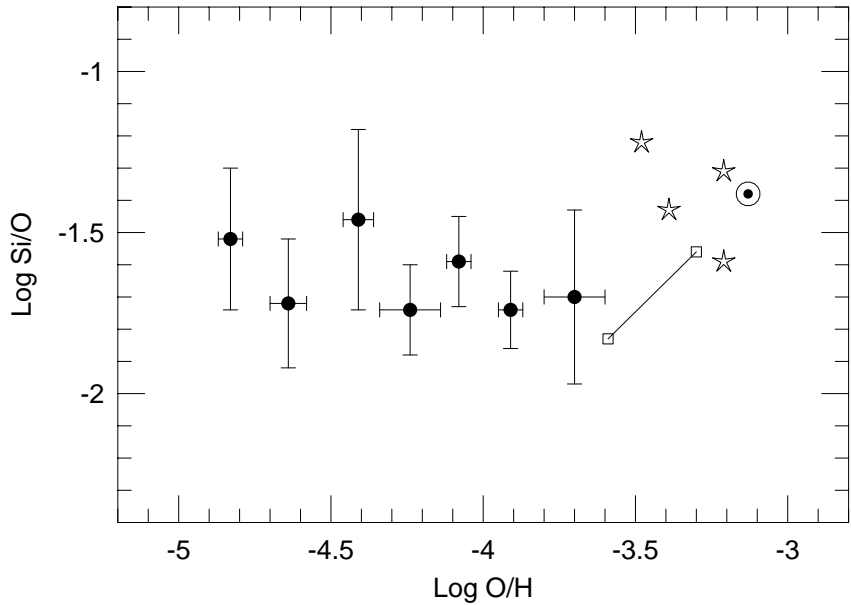

Fig. 5. Si/O vs. O/H from HST spectroscopy of H II galaxies (Garnett et al. 1995). Symbols are the same as in Figure 4.

Nevertheless, abundances for C, O, Ne, and S in the most metal-poor irregular galaxies appear to be consistent with nucleosynthesis from massive stars, with little or no contamination from long-lived intermediate mass stars. Only N seems to deviate from this picture, but primary N is probably produced in the most massive AGB stars, which have lifetimes < 50 Myr. Therefore these galaxies seem to have experienced only one or perhaps two stellar generations. Could I Zw 18 have been contaminated by another recent starburst? There is a third, less celebrated component (C) of I Zw 18, some 24″ to the NW (see Dufour & Hester 1989). We have *HST* imaging of both the main body and component C (Dufour, this volume); preliminary color-magnitude diagrams for all components show that component C is older than both the SE and NW components. Could this object have been the source of the heavy elements in the I Zw 18 main body today? A simple calculation makes the idea doubtful. The total H I mass of I Zw 18 is ≈ $7 \times 10^7$ $M_\odot$. To contaminate this much gas to the observed O/H = $1.5 \times 10^{-5}$ (Skillman & Kennicutt 1993) requires about 16,000 $M_\odot$ of oxygen, or roughly the total output from about 1000 O stars. Furthermore, the separation between C and the main body is about 1200 pc at a distance of 10 Mpc. To cross this distance in < 25 Myr requires speeds > 50 km/s, and geometric dilution of SN ejecta would require many more O stars (unless the ejecta were gravitationally focused toward the main body). Given these constraints, significant contamination from component C seems unlikely. (However, if I Zw 18 is closer than 10 Mpc – and we have only the redshift as a distance indicator – some of these problems are reduced: the separations become smaller, and the stars become fainter and older.)

*Starbursts with Coronal Winds?*

If extreme dwarf irregulars are not primordial galaxies, then it is likely that they are starbursting. Discrete starbursting behavior might be expected to have a significant effect on the chemical evolution of a dwarf galaxy. In particular, it has been suggested that episodic starbursting may affect abundance ratios of such elements, as N, Fe, and C, which come from longer lived intermediate mass stars, compared to O (e. g., Clayton & Pantelaki 1993, Garnett 1990, Gilmore & Wyse 1991). Such an effect might manifest itself as increased scatter in N/O (or C/O or Fe/O) at the lowest metallicities, but as yet there are too few accurate measurements of these ratios in very metal-poor galaxies to confirm or rule out this hypothesis.

It has also been recognized that the power from correlated Type II SNe over the short lifetime of a starburst, combined with the low gravities of dwarf galaxies, could drive a outflow which might carry away matter from a low-mass galaxy. Such outflows can reduce the effective yield of metals compared to the simple chemical evolution model (Matteucci & Chiosi 1983, Edmunds 1990). This is an attractive way to explain the low effective yields measured in dwarf irregulars, and may also explain the mass-metallicity relation for dwarf galaxies (Skillman et al. 1989a).

If the SN ejecta couple inefficiently with the ambient ISM, and instead break out through the SN-driven shell (through Rayleigh-Taylor instabilities), one can get an *enriched* outflow which may selectively remove elements like O from the system, while elements produced in AGB stars are relatively unaffected (De Young & Gallagher 1990). Pilyugin (1992) and Marconi et al. (1994) suggested that the selective loss of O from dwarf irregular galaxies could explain why the $\Delta Y - \Delta O$ relation is steeper than predicted by stellar nucleosynthesis. (However, see Traat 1995; and one can also invoke stronger stellar mass to explain this as well.) Evidence is growing for high-velocity gas flows associated with active dwarf starbursting galaxies (Meurer et al. 1992; Marlowe et al. 1995), and it is suggested that that these outflows are energetic enough to escape the galaxies. On the other hand, evidence for actual breakout of gas from dwarf galaxies has been elusive; and it should be noted that the mass-metallicity relation is continuous from the dwarf irregulars through the most massive galaxies (Zaritsky et al. 1994) – can galactic winds affect the chemical evolution of all galaxies, or is there a conspiracy (see also Skillman, this volume)?

It is clear there are still many questions to be answered. The next few years will be a good time to be studying irregular galaxies. *HST* should continue to provide new data on abundances from spectroscopy of H II regions, particularly when the STIS – the Space Telescope Imaging Spectrograph – is installed in 1997. STIS will allow us to get higher s/n and spectral resolution to measure important faint lines such as N III]. High-resolution imaging and stellar photometry with WFPC2 will also allow us to obtain exciting new data on the star formation histories of nearby irregulars. ISO and (we hope) SIRTF will provide IR spectroscopy for determining element abundances with little dependence on $T_e$. The Sloan Survey should find larger numbers of very low-luminosity, metal-poor dwarfs, while the new 8-10 meter telescopes will give us the power to observe them spectroscopically. We are also seeing the growing capability to do chemo-dynamical modeling of galaxies, which will give us a better appreciation for the evolution of irregular galaxies.

I am grateful for all of the terrific discussions I have had with my collaborators and colleagues at Mex-Tex. I also enthusiastically acknowledge the support provided by NASA and Space Telescope Science Institute through the Hubble Fellowship award HF-1030.01-92A.